\documentclass[12pt]{article}
\usepackage{graphicx}

\input{epsf}

\newcommand{\nc}{\newcommand}
\nc{\beq}{\begin{equation}}
\nc{\eeq}{\end{equation}}
\nc{\beqa}{\begin{eqnarray}}
\nc{\eeqa}{\end{eqnarray}}

\def\gsim{\mathrel{\rlap{\lower4pt\hbox{\hskip1pt$\sim$}}
    \raise1pt\hbox{$>$}}}       %greater than or approx. symbol

\begin{document}

%\baselineskip 24pt

%%%%%%%%%%%%%%%%%%%%%%%%%%%%%%%%%%%%%%%%%%%%%%%%%%%%%%%%%%%%%%%%%
%%%
%%%                      TITLE PAGE
%%%
%%%%%%%%%%%%%%%%%%%%%%%%%%%%%%%%%%%%%%%%%%%%%%%%%%%%%%%%%%%%%%%%%

\title{\large{\bf Second Order Noncommutative Corrections to Gravity}}

\author{Xavier~Calmet$^1$\thanks{xcalmet@ulb.ac.be} \ and Archil~Kobakhidze$^{2}$\thanks{kobakhid@physics.unc.edu} \\
$^1$Universit\'e Libre de Bruxelles \\ 
Service de Physique Th\'eorique, CP225 \\ 
Boulevard du Triomphe  (Campus plaine)\\
B-1050 Brussels, Belgium\\
$^2$Department of Physics and Astronomy \\
University of North Carolina at Chapel Hill \\
Chapel Hill, NC 27599, USA}
\date{May, 2006}

\maketitle

\begin{abstract}
In this work, we calculate the leading order corrections to  general relativity formulated on a canonical noncommutative spacetime. These corrections appear in the second order of the expansion in theta. First order corrections can only appear in the gravity-matter
interactions. Some implications are briefly discussed.
\end{abstract}

%%%%%%%%%%%%%%%%%%%%%%%%%%%%%%%%%%%%%%%%%%%%%%%%%%%%%%%%%%%%%%%%
%%%
%%%                     INTRODUCTION
%%%
%%%%%%%%%%%%%%%%%%%%%%%%%%%%%%%%%%%%%%%%%%%%%%%%%%%%%%%%%%%%%%%%

\newpage

It is a difficult  to formulate General Relativity on noncommutative spaces and there are thus different approaches in the literature. In \cite{Chamseddine:2000si} for example  a  deformation of Einstein's gravity was studied using a  construction  based on gauging the noncommutative SO(4,1) de Sitter group and the Seiberg-Witten map with subsequent contraction to ISO(3,1). Most recently constructions of a noncommutative gravitational theory \cite{Aschieri:2005yw,Kobakhidze:2006kb} were proposed based on  a twisted Poincar\'e  algebra \cite{Wess:2004da,Chaichian:2004za}.  The main problem in formulating a theory of gravity  on noncommutative manifolds is that  it is difficult to implement symmetries such as general coordinate covariance and local Lorentz invariance and to define derivatives which are torsion-free and satisfy the metricity condition. 

Another approach has been proposed based on true physical symmetries \cite{Calmet:2005qm,Calmet:2005mc}. In that approach one restricts the noncommutative action to symmetries of the noncommutative algebra:
\begin{eqnarray} 
[ \hat x^\mu, \hat x^\nu ]=i \theta^{\mu\nu}.
\label{a}
\end{eqnarray}
(see also \cite{Calmet:2004ii} where this idea was applied to Lorentz symmetry).
Obviously, the commutator (\ref{a}) explicitly violates general coordinate covariance since $\theta^{\mu\nu}$ is constant in all reference frames. However, we can identify a subclass of general coordinate transformations,
\begin{equation}
\hat x^{\mu \prime}=\hat x^{\mu}+\hat \xi^{\mu}(\hat x),
\label{c}
\end{equation}
which are compatible with the algebra given by (\ref{a}). The hat on the function $\hat \xi(\hat x)$ indicates that it is in the enveloping algebra.  Under the change of coordinates (\ref{c}) the commutator (\ref{a}) transforms as:
\begin{eqnarray}
[\hat x^{\mu \prime}, \hat x^{\nu \prime}]&=&
\hat x^{\mu \prime}\hat x^{\nu \prime}- \hat x^{\nu \prime} \hat x^{\mu \prime} 
%% \nonumber \\
=i \theta^{\mu\nu} + [\hat x^\mu, \hat \xi^\nu] + [\hat \xi^\mu, \hat x^\nu] + {\cal O}(\hat \xi^2) 
\label{d}
\end{eqnarray} 
Requiring that $\theta^{\mu\nu}$ remains constant yields the following partial differential equations:
\begin{eqnarray}
\theta^{\mu \alpha} \hat \partial_{\alpha} \hat  \xi^\nu(\hat x) = \theta^{\nu \beta} \hat \partial_{\beta} \hat \xi^\mu(\hat x).
\end{eqnarray}
A nontrivial solution to this condition can be easily found:
\begin{equation}\label{nctransf}
\hat \xi^{\mu}(\hat x)=\theta^{\mu\nu}\hat \partial_{\nu} \hat f(\hat x),
\label{e}
\end{equation}
where $\hat f(\hat x)$ is an arbitrary field. This noncommutative general coordinate transformation corresponds to the following \nonumber transformation: $\hat \xi^{\mu}(x)=\theta^{\mu\nu} \partial_{\nu} \hat f(x)$. The Jacobian of this restricted coordinate transformations is equal to 1, meaning that the volume element is invariant: $d^{4}x^{\prime}=d^{4}x$. The version of General Relativity based on volume-preserving diffeomorphism is known as the unimodular theory of gravitation \cite{UNI}. Thus we came to the conclusion that symmetries of canonical noncommutative spacetime naturally lead to the noncommutative version of unimodular gravity.  We obtain the noncommutative field-strength of the SO(3,1) gauge symmetry:
\begin{eqnarray}
\hat R_{ab}=R_{ab} + R^{(1)}_{ab} + R^{(2)}_{ab}+ { \cal  O}(\theta^3),
\end{eqnarray}
with
\begin{eqnarray}
R^{(1)}_{ab} =\frac{1}{2} \theta^{cd} \{R_{ac},R_{bd} \} -\frac{1}{4} \theta^{cd} \{\omega_c, (\partial_d + D_d) R_{ab} \}.
\end{eqnarray}
The noncommutative Riemann tensor is then given by
\begin{eqnarray}
\hat R_{ab}(\hat x) =\frac{1}{2}\hat R_{ab}^{\ \ cd}(\hat x) \Sigma_{cd},
\end{eqnarray}
and the leading order correction in $\theta^{ab}$ is found explicitly to be:
\begin{eqnarray}
\hat R_{ab}^{(1)  cd}(x) =\frac{1}{2} \theta^{cd}  R_{ac}^{\ \ ij} R_{bd}^{\ \ kl}  d_{ijkl}^{pq} -\frac{1}{4} 
\theta^{cd} w_c^{ij} (\partial_d + D_d) R_{ab}^{\ \ kl} d_{ijkl}^{(2)pq} 
\end{eqnarray}
the coefficient $d_{ijkl}^{(2) pq}$ is defined by
\begin{eqnarray} \label{d1}
d_{ijkl}^{(1) p q}= Tr \left ( \{ \Sigma_{ij}, \Sigma_{kl} \} \Sigma^{pg}\right),
\end{eqnarray}
where trace goes over the matrix indices of the SO(3,1) generators
$\Sigma_{ij}$. The group-theoretic coefficients of eq. (\ref{d1}) are all
vanishing by virtue of antisymmetricity of the SO(3,1) generators,
$\Sigma_{ij}^{~~T}=-\Sigma_{ij}$ and cyclic properties of the trace  \cite{monkeys}. This can
be explicitly demonstated for an arbitrary representation for the
generators, e.g. $\Sigma_{ab}=\frac{i}{4}\left[\gamma_a,\gamma_b \right]$
 
 The new result of this work is the second order correction  in $\theta^{ab}$ which is given by
 \begin{eqnarray} \label{R2}
 R^{(2) \ m n}_{rs} &=&\frac{1}{32} \theta^{ij} \theta^{kl}  ( 
    2 w^{\ ab}_i \partial_j w_k^{\ cd}  \partial_l R_{rs}^{\ \ e f}  d^{(2) m n}_{abcdef}
 - 2 R_{ik}^{\ \ ab} w_j^{\ cd} \partial_l R_{rs}^{\ \ e f}  d^{(2) m n}_{abcdef}   
 \\ \nonumber 
 && 
 - 2 \partial_l \partial_i R_{rs}^{\ \ ab} w_j^{\ cd} w_k^{\ e f} d^{(2) m n}_{abcdef}
 + 2 i w_k^{\ ab} \partial_l w_i^{\ cd} R_{rs}^{\ \ e f} w_j^{\ g h} d^{(3) m n}_{abcdefgh}  
  \\ \nonumber 
 &&
 - 4 w_k^{\ ef} \partial_l ( R_{ri}^{\ \ ab}  R_{sj}^{\ \ cd})   d^{(2) m n}_{abcdef}
 - 4 i \partial_i w_k^{\ ab} \partial_j \partial_l R_{rs}^{\ \ cd}  d^{(4) m n}_{abcd}
     \\ \nonumber 
 &&+ i w_i^{\ ab} \partial_j w_k^{\ cd} R_{rs}^{\ ef} w_l^{\ g h} d^{(5) m n}_{abcdefgh}
 -i R_{ik}^{\ ab } w_j^{\ cd} R_{rs}^{\ \ ef} w_l^{gh} d^{(5) m n}_{abcdefgh}
 \\ \nonumber &&
 -2 i w_k^{\ ab} R_{ri}^{\ \ cd} R_{sj}^{\ \ ef} w_l^{\ gh} d^{(5) m n}_{abcdefgh}
 + i w_k^{\ ab} R_{rs}^{\ \ cd} w_i^{\ ef} \partial_j w_l^{\ gh} d^{(6) m n}_{abcdefgh}
 \\ \nonumber &&
 - i w_k^{\ ab} R_{rs}^{\ \ cd} R_{il}^{\ \ ef} w_j^{\ gh}  d^{(6) m n}_{abcdefgh}
+ 2 \partial_i w_k^{\ ab} \partial_j R_{rs}^{\ \ cd} w_l^{\ ef} d^{(2) m n}_{abcdef}
  \\ \nonumber &&
  + 2 \partial_i (w_k^{\ ab} R_{rs}^{\ \ cd} ) \partial_j w_l^{\ ef} d^{(7) m n}_{abcdef}
  + 2 \partial_i R_{rk}^{\ \ ab} w_j^{\ cd} R_{sl}^{\ \ ef}  d^{(2) m n}_{abcdef}
  \\ \nonumber &&
 - 2 i w_i^{\ ab} R_{rk}^{\ \ cd} w_j^{\ ef} R_{s l}^{\ \ gh}  d^{(3) m n}_{ghabcdef}
 - 2 R_{rk}^{\ \ ab} R_i^{ \ cd} \partial_j R_{sl}^{\ \ ef} d^{(2) m n}_{efabcd}
   \\ \nonumber &&
   + R_{rk}^{\ \ ab} R_{is}^{\ \ cd} R_{jl}^{\ \  ef}  d^{(2) m n}_{efabcd}
   + 4 i \partial_i R_{rk}^{\ \ ab} \partial_j R_{sl}^{\ \ cd}  d^{(4) m n}_{abcd}
 \\ \nonumber &&
 + 2 w^{ab}_k \partial_l ( w_i^{\ cd} \partial_j R_{rs}^{\ \ ef}   d^{(2) m n}_{abcdef}
 + i w_k^{\ ab} w_i^{\ cd} \partial_j R_{rs}^{\ \ ef} w_l^{\ g h}  d^{(5) m n}_{abcdefgh}
 \\ \nonumber &&
-i w_k^{\ ab} \partial_i R_{rs}^{\ \ cd} w_j^{ e f } w_l^{\ gh} d^{(5) m n}_{abcdefgh}
- w_k^{\ ab} w_i^{ cd} R_{rs}^{ \ \ ef} w_j^{\ gh} w_l^{\ pq} d^{(8) m n}_{abcdefghpq}
 )
 \end{eqnarray}
 using the result obtained for a generic noncommutative gauge theory in \cite{Moller:2004qq} and where the coefficients $d^{(i)}$ are defined by:
 \begin{eqnarray}
d_{abcdef}^{(2) mn}&=& Tr \left (   \{\{ \Sigma_{ab}, \Sigma_{cd} \}, \Sigma_{ef} \} \Sigma^{mn}\right), \\
d_{abcdefgh}^{(3) mn}&=& Tr \left (  
 \{ \Sigma_{ab}, \{[\Sigma_{cd} , \Sigma_{ef}],\Sigma_{gh} \} \}\Sigma^{mn}\right), \\
 d_{abcd}^{(4) mn}&=& Tr \left ( [ \Sigma_{ab}, \Sigma_{cd}] \Sigma^{mn}\right), \\
 d_{abcdefgh}^{(5) mn}&=& Tr \left (  
 \{ [ \{ \Sigma_{ab}, \Sigma_{cd}\} , \Sigma_{ef}],\Sigma_{gh}  \}\Sigma^{mn}\right), \\
d_{abcdefgh}^{(6) mn}&=& Tr \left (  
 \{ [  \Sigma_{ab}, \Sigma_{cd}] , \{ \Sigma_{ef},\Sigma_{gh} \} \} \Sigma^{mn}\right), \\
 d_{abcdef}^{(7) mn}&=& Tr \left (  
 [ [  \Sigma_{ab}, \Sigma_{cd}] ,  \Sigma_{ef}] \Sigma^{mn}\right), \\
 d_{abcdefghpq}^{(8) mn}&=& Tr \left (  
 \{ [  \Sigma_{ab}, \{ [\Sigma_{cd} , \Sigma_{ef}],\Sigma_{gh} \} ], \Sigma_{pq}\} \Sigma^{mn}\right). 
\end{eqnarray}
 This coefficients are easily calculable using a specific representation, e.g. spinorial representation, for the matrices $\Sigma_{ab}$ and a computer algebra program such as Mathematica with the routine TRACER \cite{Jamin:1991dp}. We give  explicit expressions for these traces  in the appendix.
The noncommutative action is then given by
\begin{eqnarray} \label{NCaction}
S&=&\int  d^4 x \frac{1}{2 \kappa^2} \hat R(\hat x) 
=
\int  d^4 x \frac{1}{2 \kappa^2} \left (R(x)+ R^{(2)}(x)\right)+{\cal O}( \theta^3).
\end{eqnarray}
This equation is an action for the noncommutative version of the unimodular theory of gravitation. The unimodular theory is known \cite{UNI} to be classically equivalent to Einstein's General Relativity with a cosmological constant and it can be put in the form
\begin{eqnarray} \label{NCaction2}
S_{NC}= \frac{-1}{16 \pi G} \int d^4x \sqrt{-g} R(g^{\mu\nu}) + {\cal O}(\theta),
\end{eqnarray}
where $R(g^{\mu\nu})$ is the usual Ricci scalar and $g$ is the determinant of the metric. If we restrict ourselves to the transformations (\ref{nctransf}), the determinant of the metric is always equal to minus one, the term $\sqrt{-g}$ in the action is thus trivial. However, as mentioned previously, we recover full general coordinate invariance in the limit $\theta^{\mu\nu}$ to zero and it is thus important to write this term explicitly to study the symmetries of the action. In order to obtain the equations of motion corresponding to this action, we need to consider variations of (\ref{NCaction2}) that preserve 
$g=\mbox{det} g^{\mu\nu}=-1$, i.e. not all the components of $g_{\mu\nu}$ are independent. One thus introduces a new variable  $\tilde{g}_{\mu\nu}=g^\frac{1}{4}  g_{\mu\nu}$, which has explicitly a determinant equal to one.
The field equations are then
\begin{eqnarray} \label{eqofmo}
R^{\mu\nu}-\frac{1}{4} g^{\mu\nu}R+{\cal O}(\theta)=0.
\end{eqnarray}
As done in  e.g. \cite{UNI} we can use the Bianchi identities for $R$ and find:
\begin{eqnarray} 
R_{;\mu}=0
\end{eqnarray}
which can be integrated easily and give $R=\Lambda$, where $\Lambda$ is an integration constant. It can then be shown that the differential equations (\ref{eqofmo})  imply
\begin{eqnarray} \label{einstein}
R^{\mu\nu}-\frac{1}{2} g^{\mu\nu}R-\Lambda g^{\mu\nu}+{\cal O}(\theta)=0,
\end{eqnarray}
i.e. Einstein's equations of General Relativity with a cosmological constant $\Lambda$ that appears as an integration constant.  Because any solution of Einstein's equations with a cosmological constant can, at least over any topologically $R^4$ open subset of spacetime, be written in a coordinate system with  $g=-1$, the physical content of unimodular gravity is identical at the classical level to that of Einstein's gravity with some cosmological constant \cite{UNI}.
 
 The form of the ${\cal O}(\theta^2)$ corrections in eq. (\ref{R2}) suggests that in the linearized approximation, gravity is not affected by spacetime noncommutativity. Note also that in the full gravity-matter
action the dominant ${\cal O}(\theta)$ will generally be present  in the matter Lagrangian, that in turn could affect the solutions for the metric in this order. It would be very interesting to study cosmological
perturbations in the above setting.

%%%%%%%%%%%%%%%%%%%%%%%%%%%%%%%%%%%%%%%%%%%%%%%%%%%%%%%%%%%%%%%%%
%%%
%%%                   ACKNOWLEDGMENTS
%%%
%%%%%%%%%%%%%%%%%%%%%%%%%%%%%%%%%%%%%%%%%%%%%%%%%%%%%%%%%%%%%%%%%
\subsection*{Acknowledgments}
\noindent 
The work of X.C. was supported in part by the IISN and the Belgian science policy office (IAP V/27). 
%%%%%%%%%%%%%%%%%%%%%%%%%%%%%%%%%%%%%%%%%%%%%%%%%%%%%%%%%%%%%%%%%
%%%
%%%                     BIBLIOGRAPHY
%%%%%%%%%%%%%%%%%%%%%%%%%%%%%%%%%%%%%%%%%%%%%%%%%%%%%%%%%%%%%%%%%%%%

\bigskip

%\newpage
%\vskip .75 in

\section*{Appendix}
\begin{eqnarray}
d_{abcdefmn}^{(2)}&=& (\eta_{a f} \eta_{b n} \eta_{c m} \eta_{d e} - \eta_{a f} \eta_{b m} \eta_{c n} \eta_{d e}  - \eta_{ a e} \eta_{b n} \eta_{c m} \eta_{d f} + \eta_{a e} \eta_{b m} \eta_{c n} \eta_{d f} 
  \\ \nonumber &&
 - \eta_{
      a f} \eta_{b n} \eta_{c e} \eta_{d m} + \eta_{a e} \eta_{b n} \eta_{c f} \eta_{d m} 
+ \eta_{
      a f} \eta_{b e} \eta_{c n} \eta_{d m} - \eta_{a e} \eta_{b f} \eta_{c n} \eta_{d m} + 
       \\ \nonumber &&
\eta_{
      a n} (\eta_{b m} (\eta_{c f} \eta_{d e} - \eta_{c e} \eta_{d f}) + \eta_{b f} (\eta_{c e} 
      \eta_{d m} - \eta_{c m} \eta_{d e}) + \eta_{b e} (\eta_{c m} \eta_{d f} 
             - \eta_{c f} \eta_{
      d m})) 
        \\ \nonumber &&
      + \eta_{a f} \eta_{b m} \eta_{c e} \eta_{d n} 
- \eta_{a e} \eta_{b m} \eta_{c f} 
      \eta_{d n} - \eta_{a f} \eta_{b e} \eta_{c m} \eta_{d n} + \eta_{a e} \eta_{b f} \eta_{c m} 
      \eta_{d n} + \eta_{a m} (\eta_{b n} (\eta_{c e} \eta_{d f} 
      - \eta_{c f} \eta_{d e}) 
       \\ \nonumber &&
      + \eta_{b 
                  f} (\eta_{c n} \eta_{d e} - \eta_{c e} \eta_{d n}) + 
                            \eta_{b e} (\eta_{c f} \eta_{d n} - \eta_{c n} \eta_{d f})) + \eta_{a d} 
                  \eta_{b c} \eta_{e n} \eta_{f m} - \eta_{a c} \eta_{
            b d} \eta_{e n} \eta_{f m} 
                           \\ \nonumber &&
- \eta_{a d} \eta_{b c} \eta_{e m} \eta_{f 
                  n} + \eta_{a c} \eta_{b d} \eta_{e m} \eta_{f n})
\end{eqnarray}
\begin{eqnarray}
d_{abcdefghmn}^{(3)}&=& - i (\eta_{a h} \eta_{b g} \eta_{c n} \eta_{d f} \eta_{e m} - \eta_{a g} \eta_{b h} 
\eta_{c n} \eta_{d f} \eta_{e m}
   -
 \eta_{a h} \eta_{b g} \eta_{c f} \eta_{d n} \eta_{e m} + \\ \nonumber &&  \eta_{a g} \eta_{b h} \eta_{c f} \eta_{d n} \eta_{e m} +
 \eta_{a 
h} \eta_{b d} \eta_{
      c f} \eta_{g n} \eta_{e m} - \\ \nonumber && \eta_{a d} \eta_{b h} \eta_{c f} \eta_{g n} \eta_{e m} - \eta_{a 
h} \eta_{b c} \eta_{
      d f} \eta_{g n} \eta_{e m} + \eta_{a c} \eta_{b h} \eta_{d f} \eta_{g n} \eta_{e m} -
        \\ \nonumber &&
         \eta_{a h} \eta_{b g} \eta_{c m} \eta_{d f} \eta_{e n} + \\ \nonumber && \eta_{a g} \eta_{b h} \eta_{c m} \eta_{d f} \eta_{e n} +   \eta_{a 
h} \eta_{b g} \eta_{
      c f} \eta_{d m} \eta_{e n} -
      \eta_{a g} \eta_{b h} \eta_{c f} \eta_{d m} \eta_{e n} + \\ \nonumber &&
          \eta_{a n} \eta_{b m} (
      \eta_{c f} (\eta_{d g} \eta_{e h}
             - \eta_{d h} \eta_{e g})
 + \eta_{c h} (\eta_{d f} \eta_{e g} 
- \eta_{d e} \eta_{f 
      g}) \\ \nonumber &&
      + \eta_{c g} (\eta_{d e} \eta_{f h} - \eta_{d f} \eta_{e h}) +  \\ \nonumber && \eta_{c e} (\eta_{d h} 
\eta_{f g} - \eta_{d 
      g} \eta_{f h}))  + \eta_{a m} \eta_{b n} (\eta_{c f} (\eta_{d h} \eta_{e g} - \eta_{d g} \eta_{e 
h}) 
 \\ \nonumber &&
+ \eta_{c 
      h} (\eta_{d e} \eta_{f g} - \eta_{d f} \eta_{e g})
      + \eta_{c g} (\eta_{d f} \eta_{e h} - 
\eta_{d e} \eta_{f 
      h}) + \eta_{c e} (\eta_{d g} \eta_{f h} - \eta_{d h} \eta_{f g})) - \\ \nonumber && \eta_{a h} \eta_{b g} 
\eta_{c n} \eta_{d e} \eta_{f m} + \eta_{a g} \eta_{b 
      h} \eta_{c n} \eta_{d e} \eta_{f m} 
      \\ \nonumber &&
      + \eta_{a h} \eta_{b g} \eta_{c e} \eta_{d n} \eta_{f m} - 
\eta_{a g} \eta_{b h} \eta_{c e} \eta_{d 
      n} \eta_{f m} + \eta_{a h} \eta_{b g} \eta_{c m} \eta_{d e} \eta_{f n} -  \\ \nonumber &&\eta_{a g} \eta_{b h} 
\eta_{c m} \eta_{d 
      e} \eta_{f n} - \eta_{a h} \eta_{b g} \eta_{c e} \eta_{d m} \eta_{f n} + \eta_{a g} \eta_{b h} 
\eta_{c e} \eta_{d 
      m} \eta_{f n} +
       \\ \nonumber &&
        \eta_{a h} \eta_{b f} \eta_{c n} \eta_{d e} \eta_{g m} - \eta_{a f} \eta_{b h} 
\eta_{c n} \eta_{d 
      e} \eta_{g m} - \eta_{a h} \eta_{b e} \eta_{c n} \eta_{d f} \eta_{g m} + \\ \nonumber &&
       \eta_{a e} \eta_{b h} 
\eta_{c n} \eta_{d 
      f} \eta_{g m} - \eta_{a h} \eta_{b f} \eta_{c e} \eta_{d n} \eta_{g m} + \eta_{a f} \eta_{b h} 
\eta_{c e} \eta_{d 
      n} \eta_{g m}
      \\ \nonumber &&
       + \eta_{a h} \eta_{b e} \eta_{c f} \eta_{d n} \eta_{g m} - \eta_{a e} \eta_{b h} 
\eta_{c f} \eta_{d 
      n} \eta_{g m} - \eta_{a h} \eta_{b d} \eta_{c f} \eta_{e n} \eta_{g m} +
        \\ \nonumber && \eta_{a d} \eta_{b h} 
\eta_{c f} \eta_{e 
      n} \eta_{g m} 
        \\ \nonumber &&+ \eta_{a h} \eta_{b c} \eta_{d f} \eta_{e n} \eta_{g m} - \eta_{a c} \eta_{b h} 
\eta_{d f} \eta_{e 
      n} \eta_{g m} + \eta_{a h} \eta_{b d} \eta_{c e} \eta_{f n} \eta_{g m} 
      \\ \nonumber &&
      - \eta_{a d} \eta_{b h} 
\eta_{c e} \eta_{f 
      n} \eta_{g m} - \eta_{a h} \eta_{b c} \eta_{d e} \eta_{f n} \eta_{g 
      m} + \eta_{a c} \eta_{b h} \eta_{d e} \eta_{f n} \eta_{g m} - \eta_{a h} \eta_{b f} \eta_{c m} 
\eta_{d e} \eta_{g 
      n}
      \\ \nonumber &&
       + \eta_{a f} \eta_{b h} \eta_{c m} \eta_{d e} \eta_{g n} + \eta_{a h} \eta_{b e} \eta_{c m} 
\eta_{d f} \eta_{g n} - \\ \nonumber && \eta_{a e} \eta_{b 
      h} \eta_{c m} \eta_{d f} \eta_{g n} + \eta_{a h} \eta_{b f} \eta_{c e} \eta_{d m} \eta_{g n} - 
\eta_{a f} \eta_{b 
      h} \eta_{c e} \eta_{d m} \eta_{g n} - \eta_{a h} \eta_{b e} \eta_{c f} \eta_{d m} \eta_{g n} +  \\ \nonumber &&
\eta_{a e} \eta_{b 
      h} \eta_{c f} \eta_{d m} \eta_{g n} - \eta_{a h} \eta_{b d} \eta_{c e} \eta_{f m} \eta_{g n} + 
\eta_{a d} \eta_{b 
      h} \eta_{c e} \eta_{f m} \eta_{g n} + \eta_{a h} \eta_{b c} \eta_{d e} \eta_{f m} \eta_{g n} -  \\ \nonumber &&
\eta_{a c} \eta_{b 
      h} \eta_{d e} \eta_{f m} \eta_{g n} - \eta_{a g} \eta_{b f} \eta_{c n} \eta_{d e} \eta_{h m} + 
      \\ \nonumber &&
\eta_{a f} \eta_{b 
      g} \eta_{c n} \eta_{d e} \eta_{h m} +  \\ \nonumber && \eta_{a g} \eta_{b e} \eta_{c n} \eta_{d f} \eta_{h m} - 
\eta_{a e} \eta_{b 
      g} \eta_{c n} \eta_{d f} \eta_{h m} + \eta_{a g} \eta_{b f} \eta_{c e} \eta_{d n} \eta_{h m} -  \\ \nonumber &&
\eta_{a f} \eta_{b 
      g} \eta_{c e} \eta_{d n} \eta_{h m} - \eta_{a g} \eta_{b e} \eta_{c f} \eta_{d n} \eta_{h m} + 
      \\ \nonumber &&
\eta_{a e} \eta_{b g} \eta_{c f} \eta_{d n} \eta_{h m} + \eta_{a g} \eta_{b d} \eta_{c f}
       \eta_{e n} \eta_{h m} - \eta_{a d} \eta_{b g} \eta_{c f} \eta_{e n} \eta_{h m} -  \\ \nonumber &&\eta_{a g} \eta_{
                        b c} \eta_{d f} \eta_{e n} \eta_{h m} + \eta_{a c} \eta_{
                  b g} \eta_{d f} \eta_{e n} \eta_{h m} - \eta_{a g} \eta_{b d} \eta_{c e}
                         \eta_{f n} \eta_{h m} +  \\ \nonumber && \eta_{a d} \eta_{b g} \eta_{
                  c e} \eta_{f n} \eta_{h m} + \\ \nonumber && \eta_{a g} \eta_{b c}
                         \eta_{d e} \eta_{f n} \eta_{h m} - \eta_{a c} \eta_{b g} \eta_{d e} 
\eta_{f n} \eta_{h m} + \eta_{a f} \eta_{b d} \eta_{c e} \eta_{g n} \eta_{h 
      m} - \eta_{a d} \eta_{b f} \eta_{c e} \eta_{g n} \eta_{h m} - \\ \nonumber && \eta_{a e} \eta_{b d} \eta_{c f} 
\eta_{g n} \eta_{h 
      m} + \eta_{a d} \eta_{b e} \eta_{c f} \eta_{g n} \eta_{h m} - \\ \nonumber && \eta_{a f} \eta_{b c} \eta_{d e} 
\eta_{g n} \eta_{h 
      m} + \eta_{a c} \eta_{b f} \eta_{d e} \eta_{g n} \eta_{h m} + \\ \nonumber && \eta_{a e} \eta_{b c} \eta_{d f} 
\eta_{g n} \eta_{h 
      m} - \eta_{a c} \eta_{b e} \eta_{d f} \eta_{g n} \eta_{h m} +  \\ \nonumber && (-\eta_{b g} (\eta_{a f} (\eta_{c m} \eta_{d e} - 
      \eta_{c e} \eta_{d m}) + \eta_{a e} (\eta_{c f} \eta_{d m} - \eta_{c m} \eta_{d f})  \\ \nonumber && - (\eta_{a 
d} \eta_{c f} - 
      \eta_{a c} \eta_{d f}) \eta_{e m} + (\eta_{a d} \eta_{c e} - \\ \nonumber && \eta_{a c} \eta_{d e}) \eta_{f m}) 
+ \eta_{a g} (
      \eta_{b f} (\eta_{c m} \eta_{d e} -  \\ \nonumber && \eta_{c e} \eta_{d m}) + \eta_{b e} (\eta_{c f} \eta_{d m} 
- \eta_{c m} \eta_{d 
      f}) - (\eta_{b d} \eta_{c f} - \eta_{b c} \eta_{d f}) \eta_{e m} + \\ \nonumber && (\eta_{b d} \eta_{c e} - 
\eta_{b c} \eta_{d 
      e}) \eta_{f m}) - (\eta_{a d} (\eta_{b e} \eta_{c f} - \eta_{b f} \eta_{c e}) + \\ \nonumber && \eta_{a f} (
                        \eta_{b d} \eta_{c e} - \eta_{b c} \eta_{d e}) + \eta_{a 
                  e} (\eta_{b c} \eta_{d f} - \eta_{b d} \eta_{c f}) +  \eta_{a c} (\eta_{b 
                        f} \eta_{d e} - \eta_{b e} \eta_{d f})) \eta_{g m}) \eta_{h n})
\end{eqnarray}
\begin{eqnarray}
d_{abcdmn}^{(4)}&=& i (-\eta_{a m} \eta_{b d} \eta_{c n} + \eta_{a d} (\eta_{b m} \eta_{c n} - 
\eta_{b n} \eta_{c m}) \\ && \nonumber + \eta_{a c} \eta_{b n} \eta_{d m} + \eta_{a n} (\eta_{b d} \eta_{c m} -
             \eta_{b c} \eta_{d m}) + \eta_{a m} \eta_{b c}
      \eta_{d n} - \eta_{a c} \eta_{b m} \eta_{d n})
\end{eqnarray}

\begin{eqnarray}
d_{abcdefghmn}^{(5)}&=&- i (\eta_{a h} \eta_{b n} \eta_{c m} \eta_{d f} \eta_{e g} - \eta_{a h} \eta_{b m} 
\eta_{c n} \eta_{d f} \eta_{e g} - \eta_{a f} \eta_{b n} \eta_{
      c m} \eta_{d h} \eta_{e g} \\ \nonumber &&   + \eta_{a f} \eta_{b m} \eta_{c n} \eta_{d h} \eta_{e g} - \eta_{a 
h} \eta_{b n} \eta_{
      c f} \eta_{d m} \eta_{e g} +  \\  \nonumber && \eta_{a f} \eta_{b n} \eta_{c h} \eta_{d m} \eta_{e g} +   
    \\  \nonumber &&  \eta_{a h} \eta_{b f} \eta_{ c n} \eta_{d m} \eta_{e g} - \eta_{a f} \eta_{b h} \eta_{c n} \eta_{d m} \eta_{e g} +  \\ \nonumber && \eta_{a \
h} \eta_{b m} \eta_{
      c f} \eta_{d n} \eta_{e g} - \eta_{a f} \eta_{b m} \eta_{c h} \eta_{d n} \eta_{e g} -  \\ \nonumber && \eta_{a 
h} \eta_{b f} \eta_{
      c m} \eta_{d n} \eta_{e g} + \eta_{a f} \eta_{b h} \eta_{c m} \eta_{d n} \eta_{e g} -  \\ \nonumber && \eta_{a 
g} \eta_{b n} \eta_{
      c m} \eta_{d f} \eta_{e h} + \eta_{a g} \eta_{b m} \eta_{c n} \eta_{d f} \eta_{e h} +  \\  \nonumber && \eta_{a 
f} \eta_{b n} \eta_{
      c m} \eta_{d g} \eta_{e h} - \eta_{a f} \eta_{b m} \eta_{c n} \eta_{d g} \eta_{e h} + \eta_{a 
g} \eta_{b n} \eta_{
      c f} \eta_{d m} \eta_{e h} - \eta_{a f} \eta_{b n} \eta_{c g} \eta_{d m} \eta_{e h} -  \\  \nonumber && \eta_{a 
g} \eta_{b f} \eta_{
      c n} \eta_{d m} \eta_{e h} + \eta_{a f} \eta_{b g} \eta_{c n} \eta_{d m} \eta_{e h} - \eta_{a 
g} \eta_{b m} \eta_{
      c f} \eta_{d n} \eta_{e h} + \eta_{a f} \eta_{b m} \eta_{c g} \eta_{d n} \eta_{e h} +  \\  \nonumber && \eta_{a 
g} \eta_{b f} \eta_{c m} \eta_{d n} \eta_{e h} - \eta_{a f} \eta_{b g} \eta_{c m} \eta_{d n} \eta_{e h} -
       \eta_{a h} \eta_{b n} \eta_{c g} \eta_{d f} \eta_{e m} +\\  \nonumber &&  \eta_{a g} \eta_{b n} \eta_{c h} \eta_{d 
f} \eta_{e m} + \eta_{
      a h} \eta_{b g} \eta_{c n} \eta_{d f} \eta_{e m} - \eta_{a g} \eta_{b h} \eta_{c n} \eta_{d f} 
\eta_{e m} + \eta_{
      a h} \eta_{b n} \eta_{c f} \eta_{d g} \eta_{e m} - \\  \nonumber && \eta_{a f} \eta_{b n} \eta_{c h} \eta_{d g} 
\eta_{e m} - \eta_{
      a h} \eta_{b f} \eta_{c n} \eta_{d g} \eta_{e m} +  \\  \nonumber && \eta_{a f} \eta_{b h} \eta_{c n} \eta_{d g} 
\eta_{e m} - \eta_{
      a g} \eta_{b n} \eta_{c f} \eta_{d h} \eta_{e m} + \\  \nonumber &&\eta_{a f} \eta_{b n} \eta_{c g} \eta_{d h} 
\eta_{e m} + \eta_{
      a g} \eta_{b f} \eta_{c n} \eta_{d h} \eta_{e m} - \\  \nonumber &&\eta_{a f} \eta_{b g} \eta_{c n} \eta_{d h} 
\eta_{e m} - \eta_{
      a h} \eta_{b g} \eta_{c f} \eta_{d n} \eta_{e m} + \eta_{a g} \eta_{b h} \eta_{c f} \eta_{d n} 
\eta_{e m} + \eta_{
      a h} \eta_{b f} \eta_{c g} \eta_{d n} \eta_{e m} -  \\  \nonumber && \eta_{a f} \eta_{b h} \eta_{c g} \eta_{d n} 
\eta_{e m} - \eta_{
      a g} \eta_{b f} \eta_{c h} \eta_{d n} \eta_{e m} + \\  \nonumber && \eta_{a f} \eta_{b g} \eta_{c h} \eta_{d n} 
\eta_{e m} + \eta_{
      a h} \eta_{b m} \eta_{c g} \eta_{d f} \eta_{e n} - \eta_{a g} \eta_{b m} \eta_{c h} \eta_{d f} 
\eta_{e n} - \\  \nonumber && \eta_{a h} \eta_{b g} \eta_{c m} \eta_{d f} \eta_{e n} + \eta_{a g} \eta_{b 
      h} \eta_{c m} \eta_{d f} \eta_{e n} - \eta_{a h} \eta_{b m} \eta_{c f} \eta_{d g} \eta_{e n} +   \\  \nonumber &&
\eta_{a f} \eta_{b 
      m} \eta_{c h} \eta_{d g} \eta_{e n} + \eta_{a h} \eta_{b f} \eta_{c m} \eta_{d g} \eta_{e n} - \\  \nonumber &&
\eta_{a f} \eta_{b h} \eta_{c m} \eta_{d 
      g} \eta_{e n} + \eta_{a g} \eta_{b m} \eta_{c f} \eta_{d h} \eta_{e n} - \\  \nonumber && \eta_{a f} \eta_{b m} 
\eta_{c g} \eta_{d 
      h} \eta_{e n} - \eta_{a g} \eta_{b f} \eta_{c m} \eta_{d h} \eta_{e n} \\  \nonumber &&+ \eta_{a f} \eta_{b g} 
\eta_{c m} \eta_{d 
      h} \eta_{e n} + \eta_{a h} \eta_{b g} \eta_{c f} \eta_{d m} \eta_{e n} - \\  \nonumber && \eta_{a g} \eta_{b h} 
\eta_{c f} \eta_{d 
      m} \eta_{e n} - \eta_{a h} \eta_{b f} \eta_{c g} \eta_{d m} \eta_{e n} +   \\  \nonumber && \eta_{a f} \eta_{b h} 
\eta_{c g} \eta_{d 
      m} \eta_{e n} + \eta_{a g} \eta_{b f} \eta_{c h} \eta_{d m} \eta_{e n} -\\  \nonumber &&  \eta_{a f} \eta_{b g} 
\eta_{c h} \eta_{d 
      m} \eta_{e n} -\\  \nonumber && \eta_{a h} \eta_{b n} \eta_{c m} \eta_{d e} \eta_{f g} + \eta_{a h} \eta_{b m} 
\eta_{c n} \eta_{d 
      e} \eta_{f g} + \\  \nonumber && \eta_{a e} \eta_{b n} \eta_{c m} \eta_{d h} \eta_{f g} - \\  \nonumber && \eta_{a e} \eta_{b m} 
\eta_{c n} \eta_{d 
      h} \eta_{f g} + \eta_{a h} \eta_{b n} \eta_{c e} \eta_{d m} \eta_{f g} -   \\  \nonumber && \eta_{a e} \eta_{b n} 
\eta_{c h} \eta_{d m} \eta_{f g} - \\  \nonumber && \eta_{a h} \eta_{b e} \eta_{c n} \eta_{d m} \eta_{f 
      g} + \\  \nonumber &&  \eta_{a e} \eta_{b h} \eta_{c n} \eta_{d m} \eta_{f g} - \\  \nonumber && \eta_{a h} \eta_{b m} \eta_{c e} 
\eta_{d n} \eta_{f 
      g} +\\  \nonumber &&  \eta_{a e} \eta_{b m} \eta_{c h} \eta_{d n} \eta_{f g} + \eta_{a h} \eta_{b e} \eta_{c m} 
\eta_{d n} \eta_{f 
      g} - 
      \\  \nonumber &&
      \eta_{a e} \eta_{b h} \eta_{c m} \eta_{d n} \eta_{f g} + \eta_{a g} \eta_{b n} \eta_{c m} 
\eta_{d e} \eta_{f h} - \\  \nonumber && \eta_{a g} \eta_{b 
      m} \eta_{c n} \eta_{d e} \eta_{f h} - \\  \nonumber && \eta_{a e} \eta_{b n} \eta_{c m} \eta_{d g} \eta_{f h} + 
\eta_{a e} \eta_{b 
      m} \eta_{c n} \eta_{d g} \eta_{f h} - \\  \nonumber && \eta_{a g} \eta_{b n} \eta_{c e} \eta_{d m} \eta_{f h} +  \\  \nonumber && 
\eta_{a e} \eta_{b 
      n} \eta_{c g} \eta_{d m} \eta_{f h} +\\  \nonumber &&  \eta_{a g} \eta_{b e} \eta_{c n} \eta_{d m} \eta_{f h} - \\  \nonumber &&
\eta_{a e} \eta_{b 
      g} \eta_{c n} \eta_{d m} \eta_{f h} + \\  \nonumber && \eta_{a g} \eta_{b m} \eta_{c e} \eta_{d n} \eta_{f h} - 
\eta_{a e} \eta_{b 
      m} \eta_{c g} \eta_{d n} \eta_{f h} - \eta_{a g} \eta_{b e} \eta_{c m} \eta_{d n} \eta_{f h} +   \\  \nonumber &&
\eta_{a e} \eta_{b 
      g} \eta_{c m} \eta_{d n} \eta_{f h} + \\  \nonumber && \eta_{a h} \eta_{b n} \eta_{c g} \eta_{d e} \eta_{f m} - 
\eta_{a g} \eta_{b 
      n} \eta_{c h} \eta_{d e} \eta_{f m} - \\  \nonumber && \eta_{a h} \eta_{b g} \eta_{c n} \eta_{d e} \eta_{f m} + \\  \nonumber &&
\eta_{a g} \eta_{b h} \eta_{c n} \eta_{d e} \eta_{
            f m} - \eta_{a h} \eta_{b n} \eta_{c e} \eta_{d g} \eta_{f m} + 
            \\  \nonumber &&
            \eta_{a e} \eta_{b n} 
            \eta_{c h} \eta_{d g} \eta_{f m} + \\  \nonumber && \eta_{a h} \eta_{b e} \eta_{c n} \eta_{d g} \eta_{f m} 
- \eta_{a e} \eta_{b h} \eta_{c n} \eta_{d g} \eta_{f m} +   \\  \nonumber &&
\eta_{a g} \eta_{b n} \eta_{c e} \eta_{d 
            h} \eta_{f m} - \\  \nonumber && \eta_{a e} \eta_{b n} \eta_{c g} \eta_{d h} \eta_{f m} - \eta_{a g} \eta_{
            b e} \eta_{c n} \eta_{d h} \eta_{f m} + \eta_{a e} \eta_{b g} \eta_{c n} \eta_{d h} \eta_{
            f m} + \eta_{a h} \eta_{b g} \eta_{c e} \eta_{d n} \eta_{f m} -\\  \nonumber && \eta_{a g} \eta_{b h} 
\eta_{c e} \eta_{d n} \eta_{f m} - \eta_{a h} \eta_{b e} \eta_{c g} \eta_{d n} \eta_{f 
      m} + \eta_{a e} \eta_{b h} \eta_{c g} \eta_{d n} \eta_{f m} +  \\  \nonumber &&
       \eta_{a g} \eta_{b e} \eta_{c h} \
\eta_{d n} \eta_{f 
      m} - \eta_{a e} \eta_{b g} \eta_{c h} \eta_{d n} \eta_{f m} +\\  \nonumber &&  \eta_{a n} (\eta_{b f} \eta_{c m} 
\eta_{d h} \eta_{e 
      g} - \eta_{b f} \eta_{c h} \eta_{d m} \eta_{e g} + \\  \nonumber && \eta_{b g} \eta_{c m} \eta_{d f} \eta_{e h} 
- \eta_{b f} \eta_{
      c m} \eta_{d g} \eta_{e h} - \eta_{b g} \eta_{c f} \eta_{d m} \eta_{e h} +  \\  \nonumber &&
       \eta_{b f} \eta_{c \
g} \eta_{d m} \eta_{
      e h} - \eta_{b g} \eta_{c h} \eta_{d f} \eta_{e m} +\\  \nonumber &&  \eta_{b f} \eta_{c h} \eta_{d g} \eta_{e 
m} + \eta_{b g} \eta_{c f} \eta_{d h} \eta_{e m} - \eta_{b f} \eta_{c g} \eta_{d h} \eta_{e m} - \\  \nonumber && \eta_{b 
                  e} \eta_{c m} \eta_{d h} \eta_{f g} + \eta_{b e} \eta_{c h} \eta_{d m} \eta_{f 
                        g} - \eta_{b g} \eta_{c m} \eta_{d e} \eta_{f h} + \\  \nonumber &&
                  \eta_{b e} \eta_{c m} \eta_{d g} \eta_{f h} + \\  \nonumber && \eta_{b g} \eta_{c e} \eta_{
                        d m} \eta_{f h} - \eta_{b e} \eta_{c g} \eta_{d m} 
            \eta_{f h} + \eta_{b m} (\eta_{c f} (\eta_{d g} \eta_{e h} -\\  \nonumber && \eta_{d h} \eta_{e g}) + 
\eta_{c h} (\eta_{d f} \eta_{e g} - \eta_{d e} \eta_{f g}) + \eta_{c g} (\eta_{d e} \eta_{f h} - \\  \nonumber && \eta_{d 
f} \eta_{e h}) + \eta_{c e} (\eta_{d h} \eta_{f g} - \eta_{d g} \eta_{f h})) +\\  \nonumber && (\eta_{b g} (\eta_{c h}
             \eta_{d e} - \eta_{c e} \eta_{d h}) + \eta_{b e} (\eta_{c g} \eta_{d h} - \eta_{
                        c h} \eta_{d g})) \eta_{f m} + \eta_{b h} (\eta_{c f} (\eta_{d m} \eta_{
            e g} - \eta_{d g} \eta_{e m}) +\\  \nonumber && \eta_{c m} (\eta_{d e} \eta_{f g} - \eta_{d f} \eta_{e 
g}) + \eta_{c g} (\eta_{d f} \eta_{e m} - \\  \nonumber &&\eta_{d e} \eta_{f 
                  m}) + \eta_{c e} (\eta_{d g} \eta_{f m} - \\  \nonumber && \eta_{d m} \eta_{f g}))) + (
                        \eta_{a h} (\eta_{b m} (\eta_{c e} \eta_{d g} - 
                  \eta_{c g} \eta_{d e}) +\\  \nonumber && \eta_{b g} (\eta_{c m} \eta_{d e} - \eta_{c e} 
                        \eta_{d m}) + \eta_{b e} (\eta_{c g} \eta_{d m} - 
                  \eta_{c m} \eta_{d g})) +\\  \nonumber && \eta_{a g} (\eta_{b m} (\eta_{c h} \eta_{d e} - 
                        \eta_{c e} 
            \eta_{d h}) + \eta_{b h} (\eta_{c e} \eta_{d m} - \eta_{c m} \eta_{d e}) + \eta_{b e} 
(\eta_{c m} \eta_{d h} - \eta_{c h} \eta_{d m})) +\\  \nonumber && \eta_{a 
                  e} (\eta_{b m} (\eta_{c g} \eta_{d h} - \eta_{c h} \eta_{d g}) + \eta_{b 
                        h} (\eta_{c m} \eta_{d g} \\  \nonumber &&- \eta_{c g} \eta_{d m}) +
                   \eta_{b g} (\eta_{c h} \eta_{d m} - \eta_{c 
                  m} \eta_{d h}))) \eta_{f n} + \eta_{a m} (-\eta_{b f} \eta_{c n} \eta_{d 
                        h} \eta_{e g} + \eta_{b f} \eta_{c h} \eta_{d n} \eta_{
            e g} -\\  \nonumber &&  \eta_{b g} \eta_{c n} \eta_{d f} \eta_{e h} + \eta_{b f} \eta_{c n} \eta_{d g} 
\eta_{e h} +\\  \nonumber &&  \eta_{b g} \eta_{c f} \eta_{d n} \eta_{e h} - 
                  \eta_{b f} \eta_{c g} \eta_{d n} \eta_{e h} +\\  \nonumber &&  \eta_{b g} \eta_{c h} \eta_{d 
                        f} \eta_{e n} - \eta_{b f} \eta_{c h} \eta_{d g} 
                  \eta_{e n} -\\  \nonumber && \eta_{b g} \eta_{c f} \eta_{d h} \eta_{e n} + \eta_{b f} \eta_{
                        c g} \eta_{d h} \eta_{e n} +\\  \nonumber && \eta_{b e}
             \eta_{c n} \eta_{d h} \eta_{f g} - \eta_{b e} \eta_{c h} \eta_{d n} \eta_{f g} + \eta_{b 
            g} \eta_{c n} \eta_{d e} \eta_{f h} -\\  \nonumber && \eta_{b e} \eta_{c n} \eta_{d g} \eta_{f h} - \eta_{
            b g} \eta_{c e} \eta_{d n} \eta_{f h} +\\  \nonumber && \eta_{b e} \eta_{c g} \eta_{d n} \eta_{f h} + 
\eta_{b n} (\eta_{c f} (\eta_{d h} \eta_{e g} - \eta_{d g} \eta_{e h}) + \\  \nonumber &&\eta_{c h} (\eta_{d e} \eta_{
            f g} - \eta_{d f} \eta_{e g}) + \eta_{c g} (\eta_{d f} \eta_{e h} - \eta_{d e} \eta_{f 
            h}) + \eta_{c e} (\eta_{d g} \eta_{f h} -\\  \nonumber && \eta_{d h} \eta_{f g})) + (\eta_{b g} (\eta_{
            c e} \eta_{d h} - \eta_{c h} \eta_{d e}) +\\  \nonumber && \eta_{b e} (\eta_{c h} \eta_{d g} - \eta_{c 
g} \eta_{d h})) \eta_{f n} +\\  \nonumber &&  \eta_{b h} (\eta_{c f} (\eta_{d g}
                   \eta_{e n} - \eta_{d n} \eta_{e g}) +\\  \nonumber && \eta_{c n} (\eta_{d f} \eta_{e 
                        g} - \eta_{d e} \eta_{f g}) + \eta_{c g} (\eta_{d e}
                   \eta_{f n} - \eta_{d f} \eta_{e n}) +\\  \nonumber && \eta_{c e} (\eta_{d n} \eta_{f 
                        g} - \eta_{d g} \eta_{f n}))))
\end{eqnarray}
\begin{eqnarray}
d_{abcdefghmn}^{(6)}&=& i (\eta_{a m} \eta_{b d} \eta_{c n} \eta_{e h} \eta_{f g} - \eta_{a c} \eta_{b n} 
\eta_{d m} \eta_{e h} \eta_{f g} -  \\  \nonumber && \eta_{a m} \eta_{b c} 
      \eta_{d n} \eta_{e h} \eta_{f g} + \eta_{a c} \eta_{b m} \eta_{d n} \eta_{e h} \eta_{f g} - \\  \nonumber &&
\eta_{a m} \eta_{b d} 
      \eta_{c n} \eta_{e g} \eta_{f h} + \eta_{a c} \eta_{b n} \eta_{d m} \eta_{e g} \eta_{f h} +\\  \nonumber &&
\eta_{a m} \eta_{b c} 
      \eta_{d n} \eta_{e g} \eta_{f h} - \eta_{a c} \eta_{b m} \eta_{d n} \eta_{e g} \eta_{f h} + \\  \nonumber &&
\eta_{a n} (\eta_{b d} 
      \eta_{c m} - \eta_{b c} \eta_{d m}) (\eta_{e g} \eta_{f h} - \eta_{e h} \eta_{f g}) - \\  \nonumber &&
 \eta_{a 
h} \eta_{b d} \eta_{
      c g} \eta_{e n} \eta_{f m} + \eta_{a g} \eta_{b d} \eta_{c h} \eta_{e n} \eta_{f m} +\\  \nonumber && \eta_{a 
h} \eta_{b c} \eta_{
      d g} \eta_{e n} \eta_{f m} - \eta_{a c} \eta_{b h} \eta_{d g} \eta_{e n} \eta_{f m} -\\  \nonumber && \eta_{a 
g} \eta_{b c} \eta_{
      d h} \eta_{e n} \eta_{f m} + \eta_{a c} \eta_{b g} \eta_{d h} \eta_{e n} \eta_{f m} +\\  \nonumber && \eta_{a 
h} \eta_{b d} \eta_{
      c g} \eta_{e m} \eta_{f n} - \eta_{a g} \eta_{b d} \eta_{c h} \eta_{e m} 
            \eta_{f n} -\\  \nonumber &&  \eta_{a h} \eta_{b c} \eta_{d g} \eta_{e m} \eta_{f n} + \\  \nonumber &&\eta_{a c} \eta_{b 
h} \eta_{d g} \eta_{e m} \eta_{f n} + \eta_{a g} \eta_{b c} 
      \eta_{d h} \eta_{e m} \eta_{f n} - \eta_{a c} \eta_{b g} \eta_{d h} \eta_{e m} \eta_{f n} + \\  \nonumber &&
\eta_{a h} \eta_{b d} \eta_{c f} \eta_{e n} \eta_{g m} - \eta_{a f} \eta_{b d} \eta_{c h} 
      \eta_{e n} \eta_{g m} - \eta_{a h} \eta_{b c} \eta_{d f} \eta_{e n} \eta_{g m} + \eta_{a c} 
\eta_{b h} \eta_{d f} \eta_{e n} \eta_{g m} + \\  \nonumber &&
      \eta_{a f} \eta_{b c} \eta_{d h} \eta_{e n} \eta_{g m} - \eta_{a c} \eta_{b f} \eta_{d h} \eta_{e 
n} \eta_{g m} - \\  \nonumber &&
      \eta_{a h} \eta_{b d} \eta_{c e} \eta_{f n} \eta_{g m} + \eta_{a e} \eta_{b d} \eta_{c h} \eta_{f 
n} \eta_{g m} + \\  \nonumber &&
      \eta_{a h} \eta_{b c} \eta_{d e} \eta_{f n} \eta_{g m} - \eta_{a c} \eta_{b h} \eta_{d e} \eta_{f 
n} \eta_{g m} - \\  \nonumber &&
      \eta_{a e} \eta_{b c} \eta_{d h} \eta_{f n} \eta_{g m} + \eta_{a c} \eta_{b e} \eta_{d h} \eta_{f 
n} \eta_{g m} - \\  \nonumber &&
      \eta_{a h} \eta_{b d} \eta_{c f} \eta_{e m} \eta_{g n} + \eta_{a f} \eta_{b d} \eta_{c h} \eta_{e 
m} \eta_{g n} + \\  \nonumber &&
      \eta_{a h} \eta_{b c} \eta_{d f} \eta_{e m} \eta_{g n} - \eta_{a c} \eta_{b h} \eta_{d f} \eta_{e 
m} \eta_{g n} - \\  \nonumber &&
      \eta_{a f} \eta_{b c} \eta_{d h} \eta_{e m} \eta_{g n} + \eta_{a c} \eta_{b f} \eta_{d h} \eta_{e 
m} \eta_{g n} + \\  \nonumber &&
      \eta_{a h} \eta_{b d} \eta_{c e} \eta_{f m} \eta_{g n} - \eta_{a e} \eta_{b d} \eta_{c h} \eta_{f 
m} \eta_{g n} - \\  \nonumber &&
      \eta_{a h} \eta_{b c} \eta_{d e} \eta_{f m} \eta_{g n} + \eta_{a c} \eta_{b h} \eta_{d e} \eta_{f 
m} \eta_{g n} + \\  \nonumber && \eta_{a e} \eta_{b c} \eta_{d h} \eta_{f m} \eta_{g n} - \eta_{a c} 
      \eta_{b e} \eta_{d h} \eta_{f m} \eta_{g n} - \eta_{a g} \eta_{b d} \eta_{c f} \eta_{e n} \eta_{h 
m} + \\  \nonumber && \eta_{a f} 
      \eta_{b d} \eta_{c g} \eta_{e n} \eta_{h m} + \eta_{a g} \eta_{b c} \eta_{d f} \eta_{e n} \eta_{h 
m} - \eta_{a c} \eta_{b g} \eta_{d f} 
      \eta_{e n} \eta_{h m} - \\  \nonumber && \eta_{a f} \eta_{b c} \eta_{d g} \eta_{e n} \eta_{h m} + \eta_{a c} 
\eta_{b f} \eta_{d g} 
      \eta_{e n} \eta_{h m} +\\  \nonumber && \eta_{a g} \eta_{b d} \eta_{c e} \eta_{f n} \eta_{h m} - \eta_{a e} 
\eta_{b d} \eta_{c g} 
      \eta_{f n} \eta_{h m} - \\  \nonumber &&\eta_{a g} \eta_{b c} \eta_{d e} \eta_{f n} \eta_{h m} + \eta_{a c} 
\eta_{b g} \eta_{d e} 
      \eta_{f n} \eta_{h m} +\\  \nonumber && \eta_{a e} \eta_{b c} \eta_{d g} \eta_{f n} \eta_{h m} - \eta_{a c} 
\eta_{b e} \eta_{d g} 
      \eta_{f n} \eta_{h m} -\\  \nonumber && \eta_{a f} \eta_{b d} \eta_{c e} \eta_{g n} \eta_{h m} + \eta_{a e} 
\eta_{b d} \eta_{c f} 
      \eta_{g n} \eta_{h m} + \\  \nonumber &&\eta_{a f} \eta_{b c} \eta_{d e} \eta_{g n} \eta_{h m} - \eta_{a c} 
\eta_{b f} \eta_{d e} 
      \eta_{g n} \eta_{h m} - \\  \nonumber &&\eta_{a e} \eta_{b c} \eta_{d f} \eta_{g n} \eta_{h m} + \eta_{a c} 
\eta_{b e} \eta_{d f} 
      \eta_{g n} \eta_{h m} +\\  \nonumber && (\eta_{a c} \eta_{b g} \eta_{d f} \eta_{e m} - \eta_{a c} \eta_{b f} 
\eta_{d g} \eta_{e m} +\\  \nonumber && \eta_{a e} \eta_{b d} 
            \eta_{c g} \eta_{f m} - \eta_{a c} \eta_{b g} \eta_{d e} \eta_{f m} - \eta_{a e} \eta_{b 
c} \eta_{d g} \eta_{f m} +\\  \nonumber && \eta_{a c} \eta_{b e} \eta_{d g} 
                  \eta_{f m} + \eta_{a g} (\eta_{b d} (\eta_{c f} \eta_{e m} - \eta_{c e} 
                        \eta_{f m}) + \\  \nonumber &&\eta_{b c} (\eta_{d e} \eta_{f m} - 
                  \eta_{d f} \eta_{e m})) +\\  \nonumber && (\eta_{a e} (\eta_{b c} \eta_{d f} - \eta_{b d}
                         \eta_{c f}) +\\  \nonumber && \eta_{a c} (\eta_{b f} \eta_{d e} - \eta_{b e} \eta_{d 
                        f})) \eta_{g m} + \\  \nonumber &&\eta_{a f} (\eta_{b d} (\eta_{c e} \eta_{g m} - \eta_{
            c g} \eta_{e m}) + \\  \nonumber &&\eta_{b c} (\eta_{d g} \eta_{e m} - \eta_{d e} \eta_{g m}))) 
\eta_{h n} + \eta_{a d} ((\eta_{b n} \eta_{c m} - \eta_{b m} \eta_{
                  c n}) (\eta_{e h} \eta_{f g} -\\  \nonumber && \eta_{e g} \eta_{f h}) + \eta_{b f} \eta_{
                        c h} \eta_{e n} \eta_{g m} - \eta_{b e} \eta_{c h} \eta_{f n} \eta_{g 
m} - \eta_{b f} \eta_{c h} \eta_{e m} \eta_{g n} +\\  \nonumber && \eta_{b e} \eta_{c h} \eta_{f m}
                   \eta_{g n} +\\  \nonumber && \eta_{b h} (\eta_{c g} (\eta_{e n} \eta_{f m} - \eta_{e m} \eta_{f 
n}) + \eta_{c f} (\eta_{e m} \eta_{g n} - \\  \nonumber &&\eta_{e n} \eta_{g 
            m}) + \eta_{c e} (\eta_{f n} \eta_{g m} -\\  \nonumber && \eta_{f m} \eta_{g n})) - \eta_{b f} \eta_{c \
g} \eta_{e n} \eta_{h m} + \\  \nonumber &&\eta_{b e} \eta_{c g} \eta_{f n} \eta_{h m} + \eta_{b f} \eta_{c e} \eta_{
            g n} \eta_{h m} - \\  \nonumber &&\eta_{b e} \eta_{c f} \eta_{g n} \eta_{h m} + (\eta_{b f} (\eta_{c g}
             \eta_{e m} - \\  \nonumber &&\eta_{c e} \eta_{g m}) + \eta_{b e} (\eta_{c f} \eta_{g m} - \eta_{c g} 
            \eta_{f m})) \eta_{h n} + \\  \nonumber &&\eta_{b g} (\eta_{c h} (\eta_{e m} \eta_{f n} - \eta_{e n} 
\eta_{f m}) + \eta_{c f} (\eta_{e n} \eta_{h m} - \eta_{e m} \eta_{h 
                  n}) +\\  \nonumber &&  \eta_{c e} (\eta_{f m} \eta_{h n} - \eta_{f n} \eta_{h m}))))
\end{eqnarray}

\begin{eqnarray}
d_{abcdefmn}^{(7)}&=& (-\eta_{a f} \eta_{b d} \eta_{c n} \eta_{e m} + \eta_{a c} \eta_{b n} \eta_{d f} \eta_{e m} +\\  \nonumber &&  \eta_{a 
      f} \eta_{b c} \eta_{d n} \eta_{e m} - \eta_{a c} \eta_{b f} \eta_{d n} \eta_{e m} - \eta_{a 
      m} \eta_{b d} \eta_{c f} \eta_{e n} +\\  \nonumber && \eta_{a f} \eta_{b d} \eta_{c m} \eta_{e n} + \eta_{a 
      m} \eta_{b c} \eta_{d f} \eta_{e n} -\eta_{a c} \eta_{b m} \eta_{d f} \eta_{e n} - \eta_{a 
      f} \eta_{b c} \eta_{d m} \eta_{e n} +\\  \nonumber && \eta_{a c} \eta_{b f} \eta_{d m} \eta_{e n} + \eta_{a 
      e} \eta_{b d} \eta_{c n} \eta_{f m} - \eta_{a c} \eta_{b n} \eta_{d e} \eta_{f m} - \eta_{a 
      e} \eta_{b c} \eta_{d n} \eta_{f m} +\\  \nonumber && \eta_{a c} \eta_{b e} \eta_{d n} \eta_{f m} + \eta_{a 
      n} (\eta_{b d} (\eta_{c f} \eta_{e m} - \eta_{c e} \eta_{f m}) + \eta_{b c} (\eta_{d e} 
      \eta_{f m} - \eta_{d f} \eta_{e m})) +\\  \nonumber && (\eta_{a m} (\eta_{b d} \eta_{c e} - \eta_{b c} \eta_{
      d e}) + \eta_{a e} (\eta_{b c} \eta_{d m} -  \eta_{b d} \eta_{c m}) + \eta_{a c} (\eta_{b 
      m} \eta_{d e} - \eta_{b e} \eta_{d m})) \eta_{f n} +\\  \nonumber && \eta_{a d} (\eta_{b f} (\eta_{c n} 
      \eta_{e m} - \eta_{c m} \eta_{e n}) +  \eta_{b n} (\eta_{c e} \eta_{f m} - \eta_{c f} \eta_{e 
      m}) + \eta_{b m} (\eta_{c f} \eta_{e n} -\\  \nonumber && \eta_{c e} \eta_{f n}) + \eta_{b e} (\eta_{c m}
       \eta_{f n} - \eta_{c n} \eta_{f m})))
\end{eqnarray}

\begin{eqnarray}
d_{abcdefghpqmn}^{(8)}&=& -\frac{1}{2}(\eta_{de} \epsilon_{cfgh} - \eta_{ec} \epsilon_{dfgh} - \epsilon_{edgh} \eta_{fc} + \epsilon_{ecgh} \eta_{ld})\\  \nonumber &&
( \epsilon_{bmnp} \eta_{aq} -\epsilon_{anpq} \eta_{bm} + \epsilon_{ampq} \eta_{bn} -\epsilon_{amnq} \eta_{bp} + \epsilon_{amnp} \eta_{bq}) 
\end{eqnarray}
\bigskip

\baselineskip=1.6pt
\begin{thebibliography}{99}

%\cite{Chamseddine:2000si}
\bibitem{Chamseddine:2000si}
  A.~H.~Chamseddine,
 %``Deforming Einstein's gravity,''
  Phys.\ Lett.\ B {\bf 504}, 33 (2001)
  [arXiv:hep-th/0009153].
  %%CITATION = HEP-TH 0009153;%%


%\cite{Aschieri:2005yw}
\bibitem{Aschieri:2005yw}
  P.~Aschieri, C.~Blohmann, M.~Dimitrijevic, F.~Meyer, P.~Schupp and J.~Wess,
  %``A gravity theory on noncommutative spaces,''
  Class.\ Quant.\ Grav.\  {\bf 22}, 3511 (2005)
  [arXiv:hep-th/0504183];
  %%CITATION = HEP-TH 0504183;%%
%\cite{Aschieri:2005zs}
%\bibitem{Aschieri:2005zs}
  P.~Aschieri, M.~Dimitrijevic, F.~Meyer and J.~Wess,
  %``Noncommutative geometry and gravity,''
  Class.\ Quant.\ Grav.\  {\bf 23}, 1883 (2006)
  [arXiv:hep-th/0510059].
  %%CITATION = HEP-TH 0510059;%%
  
  
%\cite{Kobakhidze:2006kb}
\bibitem{Kobakhidze:2006kb}
  A.~Kobakhidze,
  %``Theta-twisted gravity,''
  arXiv:hep-th/0603132.
  %%CITATION = HEP-TH 0603132;%%
  
%\cite{Wess:2004da}
\bibitem{Wess:2004da}
J.~Wess,
``Deformed coordinate spaces: Derivatives,'' in Lectures given at BW2003 Workshop on Mathematical, Theoretical and Phenomenological Challenges Beyond the Standard Model: Perspectives of Balkans Collaboration, Vrnjacka Banja, Serbia, 29 Aug - 2 Sep 2003, 
arXiv:hep-th/0408080.
%%CITATION = HEP-TH 0408080;%%
%\cite{Chaichian:2004za}

\bibitem{Chaichian:2004za}
%\bibitem{Chaichian:2004za}
  M.~Chaichian, P.~P.~Kulish, K.~Nishijima and A.~Tureanu,
 %``On a Lorentz-invariant interpretation of noncommutative spacetime and its implications on noncommutative QFT,''
  Phys.\ Lett.\ B {\bf 604}, 98 (2004)
  [arXiv:hep-th/0408069].
  %%CITATION = HEP-TH 0408069;%%
  
  
  %\cite{Calmet:2005qm}
\bibitem{Calmet:2005qm}
  X.~Calmet and A.~Kobakhidze,
  %``Noncommutative general relativity,''
  Phys.\ Rev.\ D {\bf 72}, 045010 (2005)
  [arXiv:hep-th/0506157].
  %%CITATION = HEP-TH 0506157;%%
  
  %\cite{Calmet:2005mc}
\bibitem{Calmet:2005mc}
  X.~Calmet,
  %``Cosmological constant and noncommutative spacetime,''
  arXiv:hep-th/0510165.
  %%CITATION = HEP-TH 0510165;%%
  
  
%\cite{Calmet:2004ii}
\bibitem{Calmet:2004ii}
 X.~Calmet, 
 %``Spacetime symmetries of noncommutative spaces,'' 
 Phys.\ Rev.\ D {\bf 71}, 085012 (2005) [arXiv:hep-th/0411147] and 
  %%CITATION = HEP-TH 0411147;%%
%\cite{Calmet:2006iy}
%\bibitem{Calmet:2006iy}
  %``Symmetries, microcausality and physics on canonical noncommutative
  %spacetime,''
  arXiv:hep-th/0605033, to appear in the proceedings of 41st Rencontres de Moriond on Electroweak Interactions and Unified Theories, La Thuile, Aosta Valley, Italy, 11-18 Mar 2006. 
  %%CITATION = HEP-TH 0605033;%%
  
  %\cite{Jamin:1991dp}
\bibitem{Jamin:1991dp}
  M.~Jamin and M.~E.~Lautenbacher,
  %``TRACER: Version 1.1: A Mathematica package for gamma algebra in arbitrary
  %dimensions,''
  Comput.\ Phys.\ Commun.\  {\bf 74} (1993) 265.
  %%CITATION = CPHCB,74,265;%%
    
  %\cite{Moller:2004qq}
\bibitem{Moller:2004qq}
  L.~Moller,
  %``Second order of the expansions of action functionals of the  noncommutative
  %standard model,''
  JHEP {\bf 0410}, 063 (2004)
  [arXiv:hep-th/0409085].
  %%CITATION = HEP-TH 0409085;%%
  
  \bibitem{UNI}
The equations of motion corresponding to this theory have first been written down  by A.~Einstein in:
A.~Einstein, Siz. \ Preuss. \ Acad. \ Scis., (1919); ``Do Gravitational Fields Play an essential Role in the Structure 
of Elementary Particle of Matter,'' by A.~Einstein {\it et al} (Dover, New York, 1952)[Eng. translation]. \\
The theory has been rediscovered in 
J.~J.~van der Bij, H.~van Dam and Y.~J.~Ng, 
%``Theory Of Gravity And The Cosmological Term: The Little Group Viewpoint,'' 
Physica {\bf 116A}, 307 (1982),
and further developed by a number of authors, see e.g., 
F.~Wilczek,
%``Foundations And Working Pictures In Microphysical Cosmology,'' 
Phys.\ Rept.\  {\bf 104}, 143 (1984); 
W.~Buchmuller and N.~Dragon, 
%``Einstein Gravity From Restricted Coordinate Invariance,''
Phys.\ Lett.\ B {\bf 207}, 292 (1988); 
M.~Henneaux and C.~Teitelboim,
%``The Cosmological Constant And General Covariance,'' 
Phys.\ Lett.\ B {\bf 222}, 195 (1989); W.~G.~Unruh,
%``A Unimodular Theory Of Canonical Quantum Gravity,'' 
Phys.\ Rev.\ D {\bf 40}, 1048 (1989).


  \bibitem{monkeys}
P. ~Mukherjee, private communication.



\end{thebibliography}
\end{document}